# Increased superconducting transition temperature of a niobium thin-film proximity-coupled to gold nanoparticles using linking organic molecules

Eran Katzir,[1] Shira Yochelis,[1] Felix Zeides,[2] Nadav Katz,[2] Yoav Kalcheim,[2] Oded Millo,[2] Gregory Leitus,[3] Yuri Myasodeyov,[4] Boris Ya. Shapiro,[5] Ron Naaman,[6] and Yossi Paltiel[1]

[1] Applied Physics Department and the Center for Nano-Science and Nano-Technology, the Hebrew University of Jerusalem, Jerusalem, 91904 Israel.
[2] Racah Institute of Physics the Hebrew University of Jerusalem, Jerusalem, 91904 Israel
[3] Chemical Support, The Weizmann Institute, Rehovot 76100, Israel
[4] Department of Physics, The Weizmann Institute of Science, Rehovot, 76100 Israel.
[5] Physics Department, Bar-Ilan University, Ramat-Gan, 52900 Israel
[6] Department of Chemical Physics, The Weizmann Institute of Science, Rehovot, 76100 Israel.

*Corresponding author: Yossi Paltiel- paltiel@cc.huji.ac.il*

## Abstract:

The superconducting critical temperature, $T_C$, of thin Nb films is significantly modified when gold nanoparticles (NPs) are chemically linked to the Nb film, with a consistent enhancement when using 3 nm long disilane linker molecules. The $T_C$ increases by up to 10% for certain linker length and NPs size. No change is observed when the nanoparticles are physisorbed with non-linking molecules. Electron tunneling spectra acquired on the linked NPs below $T_C$ typically exhibit zero-bias peaks. We attribute these results to a pairing mechanism coupling electrons in the Nb and the NPs, mediated by the organic linkers.

The proximity effect (PE) in superconductivity occurs when a superconductor (*S*) is placed in contact with a normal metal (*N*). The resulting critical temperature of the superconductor, $T_C$, is then suppressed and signs of weak superconductivity are induced in *N* [1]. The PE is well understood for 'conventional' BCS superconductors, and is known to take place via Andreev reflections. By this, superconductivity is weakened in the superconductor and superconducting correlations are induced in the normal side up to a distance where the electron and hole lose phase-coherence [1]. Hence, the PE requires good electrical contact and Fermi wave-vector matching. Nevertheless, and quite surprisingly, PEs were observed also for non-metallic systems such as Anderson insulators [2], molecules [3], and hybrid superconductor quantum-dot devices [4].

Even more intriguing, and less common, is the inverse PE, where the $T_C$ of a superconductor is increased upon attachment to a non-superconducting material. Such an effect was reported for quench-condensed ultrathin Pb film (*S*), where $T_C$ was increased upon coverage with Ag (*N*) [5] The authors attributed their findings to Coulomb repulsion screening. $T_C$ enhancement was recently observed also for *S*-*N* bilayers involving cuprate high-temperature superconductors [6].

Here we report the observation of a robust inverse PE effect for a self-assembled monolayer of gold nanoparticles (NPs), attached via organic linker molecules to Nb film [7]. The critical temperature of the Nb film increased substantially in such heterostrucures, by up to ~10%, compared to the bare film. The relative enhancement is more pronounced when the Nb films are thinner. Concomitantly, tunneling spectra acquired on the linked Au NPs, below $T_C$, typically exhibit zero-bias peaks. Only in the case of the shortest linkers, for some samples induced gaps are also observed, signifying an interface effect. It is important to note that in contrast to previous bulk planer geometries; the inverse PE measured here involves a local effect by the use of metallic nano-dots with organic molecules.

In conventional superconductivity, $T_C$ is governed by the interactions leading to electron Cooper pairing. This is in contrast to granular or under-doped high-$T_C$ superconductors, where $T_C$ is governed by phase fluctuations [8]. It is well known that the critical temperature and the gap can be enhanced by a few percent by driving the superconductor out of equilibrium [9]. When the superconductors are carefully arranged, the critical temperature depends not only on the strength of the pairing interactions but also on quasiparticle excitations. Optimizing the tunnel injector-extractor arrangement and the thickness of Al film sandwiched between two tunnel junctions in the Nb-AlO$_x$-Al- AlO$_x$-Nb configuration has led to almost 100% enhancement of $T_C$ in the Al (S) layer [10].

In the present work we investigated modifications in the superconductor properties of 150 and 50-nm-thick Nb films due to adsorption of organic molecules and Au NPs. We used two types of Nb films, one evaporated on sapphire and the other sputtered on Si wafers. The $T_C$ of the sputtered film was higher; nevertheless both samples gave the same order of $T_C$ enhancement upon the attachment of the Au NPs. The corrugations of the Nb films were of the order of 2-3 nm over an area of few μm$^2$. The films were coated with self-assembled monolayers of two types of organic linkers, to which the Au NPs were attached (see Fig. 1a), 3-methylpropane bis-trichlorosilane (di-silane), which forms a 3-nm-thick layer, and mercapto propyl silane (MPS), which forms a 0.5-nm-thick monolayer. We also studied a sample covered by ~3 nm long non-linking octadecyl trichlorosilane (OTS) molecules, on which the Au NPs were only physisorbed and thus acted merely as a spacer layer. The thicknesses of the organic films were measured using an ellipsometer as well as by XPS (see supplementary material). The Au NPs, either 10 or 5 nm in diameter, were attached to the organic molecules using the procedure described in Ref. [11].

The measured $T_C$ values of the Nb films ranged from 7.8 K to 8.5 K and the magnetic field, normal to the sample surface, could be varied up to 1T. For the sputtered films, the superconducting transition width was ~0.05 K, implying that the Nb films were quite pure. The transition was two times wider for the evaporated film. The scanning tunneling microscopy (STM) and spectroscopy measurements were performed in a home-built cryogenic STM, using Pt-Ir tips, operating in a clean He exchange-gas environment.

Figure 1(b) presents the derivative of the resistance as a function of temperature, normalized to the $T_C$ of the bare evaporated Nb film of thickness 150 nm, which was 7.8K. Data for the bare film (black curve) is shown as well as data obtained after the attachment of 5 and 10 nm diameter Au NPs with different organic linkers. The inset of Fig. 1(b) shows a SEM image of the monolayer made from the 10-nm Au NPs on top of the Nb film. All samples were cooled at zero magnetic fields, and the measurements were performed in the cooling direction using different currents. The results presented are the ones with the highest current for which no heating effect was observed.

When using the Nb film and short (MPS) organic linker, the binding of the Au NPs yields a reduction of the critical temperature. This reduction is more pronounced with the larger NPs (10 nm dia.) as compared with the small ones (5 nm dia.). These results are consistent with the expected PE and occurred also when the Au NPs were physisorbed directly on the surface of the Nb film. However, when longer linker molecules were used, a surprising increase in $T_C$ was observed, by 3% for the 10-nm Au NPs and by 2% for the 5-nm Au NPs. Similar results were measured using the sputtered Nb film.

Figure 1(c) presents the measurements on the 50nm thick sputtered Nb films coated with MPS and di-silane molecules to which 10 nm Au NPs were covalently attached. Here, an increase of about 10% in the critical temperature is measured for the MPS molecule (green curve), comparable to that measured for the di-silane molecule (red curve). For comparison results are shown for the NPs physisorbed on the non-linking OTS layer (blue curve). A negligible effect on $T_C$ is observed in this case, although the thickness of this layer is comparable to that of di-silane. The widening of the transition fits the results of the phase transition, where only organic molecules are adsorbed [12]. This observation confirms that screening effects are not involved [5], since here the OTS layer acts only as a spacer. We also find that the adsorbed monolayer by itself, with no NPs, does not affect the critical temperature consistent with Ref [12]. The increase in the critical temperature was similar for both the sputtered film with higher $T_C$ and the evaporated film of the same thickness. Since the adsorption of NPs is not uniform, it leads to local fluctuations of the critical current [13] and probably to local $T_C$ fluctuations. This effect could be used for patterning areas with high $T_C$ on the Nb film, by selective adsorption of the NPs.

The dependence of $T_C$ on the density of the adsorbed Au NPs is demonstrated in Fig. 2, where the temperature dependent resistance, $R(T)$, is presented for different NPs densities. After deposition of 10 nm dia. Au NPs on the di-silane layer, $T_C$ increased by 3% as compared with the bare Nb substrate. Removing the covalent bounded NPs is difficult, and only by applying plasma cleaning we were able to affect the organic layer. With each exposure to the plasma, the organic monolayer was damaged and consequently the NPs density was gradually reduced. Indeed, when the density of NPs was reduced, $T_C$ decreased monotonically. The plasma process effect on the bare sample was small compare to figure 2 results.

For a sample with density of approximately $6\times10^{10}$ NPs/cm$^2$, the superconducting phase transition was measured at different magnetic fields. The density of the vortices increases with the magnetic field [14] and the *Tc* of the sample decreases. At a magnetic field of about 1 T, the density of the vortices matches the density of the NPs. Around this 'matching field' the $T_C$ enhancement is reduced, as is evident by comparing the inset with the main Fig. 2. We relate the shoulder in *R(T)*, measured at 9000 G on the bare Nb film (blue line in the inset), to vortex creep mechanism [15]. The disappearance of the shoulder after the attachment of the Au NPs indicates that the NPs provide efficient pinning centers, manifesting their electrical coupling to the Nb film.

The above results suggest unique coupling occurs between the electrons in the Nb film and in the NPs, mediated by the organic linkers. Such an effect is expected to manifest itself by modification of the electronic density of states (DOS) of the NPs and/or of the Nb film. To study this effect, we performed electron tunneling spectroscopy measurements, using STM, on the various samples described above, at 4.2 K and at 15 K (below and above $T_C$). Zero bias anomalies, mostly peaks but occasionally also gaps in the DOS, were observed below $T_C$ for samples in which the NPs caused a change in $T_C$.

Figure 3 shows two STM *dI/dV* versus *V* tunneling spectra acquired at 4.2 K on a 10 nm dia. In this case the Au NPs (shown to the left of the spectra) were linked to a 50 nm thick Nb film via disilane molecule. Such tunneling spectra, which are proportional to the local DOS, were numerically derived from *I-V* curves that were acquired while momentarily disconnecting the STM feedback loop (I-V curves are presented in Fig. 4 of the supplementary material). The figure presents a pronounced zero bias peak observed below $T_C$ only, when measuring on Au NPs in samples that exhibit $T_C$ enhancement. Above $T_C$, this peak vanished and the spectra became featureless (metallic-like). Such zero-bias peaks were observed for a wide range of bias or current set-points, being nearly reproducible for a given setting (figure 3), thus excluding the possibility that they are related to single electron charging effects [16]. The STM setting did affect, however, the peak width, that ranged between 2 to 10 mV, and its height (see supplementary material). The tunneling spectra, measured over the organic layer far from Au NPs (in all samples), showed small gaps such as the one presented in the left inset to Fig. 3. Fitting such a gapped conductance spectra to the Dynes DOS formula [17], $\rho(E) = \left| \text{Re}\left( \frac{E - i\Gamma}{((E - i\Gamma)^2 - \Delta^2)^{1/2}} \right) \right|$, yielded superconducting gap values, $\Delta$, in the range 0.81-0.85 meV, somewhat smaller than the largest gaps reported for bare Nb surface, ~1.2 meV [18], and relatively large broadening parameters, $\Gamma$ ~0.32 meV, probably due to the presence of the organic molecules.

We occasionally found very shallow induced proximity gaps on NPs linked to the Nb via the short MPS molecules, along with the conductance peaks observed on most of the other NPs. Such a gap is presented in the right inset of Fig. 3 and could be fitted to $\Delta = 0.8$ meV and $\Gamma = 1.4$ meV. This may relate to our observation that the MPS molecules can lead to either enhancement or reduction in $T_C$ typical to the conventional PE. It should also be pointed out that the absence of single electron charging effects, in this sample, suggests that the chemically linked NPs are well coupled electronically to the Nb film. In samples where the NPs were physisorbed on the OTS molecules, a different behavior was observed and the tunneling spectra frequently exhibited single electron charging effects [16].

The observed $T_C$ enhancement effect is robust and does not depend (qualitatively) on the sample preparation procedure or the $T_C$ of the bare Nb film (for the same film thickness). However, the magnitude of this effect does depend on the film thickness, the NP size, and the degree of coupling between the NPs and the Nb film.

Several mechanisms were discussed in the literature that may contribute to the observed effect. Among them are enhancement of superconductivity by Anderson localization induced by pair pinning in the vicinity of the NPs [19], or suppressing the surface phonon effect that tend to reduce the $T_C$ of the bare thin Nb films [20]. Another approach is to relate our results to a mechanism suggested by Leggett et al. [21] for high $T_C$ superconductors, [21] [22] in which Coulomb interaction between neighboring $CuO_2$ plains changes the Cooper pairing Hamiltonian and thus increases the inter-plain coupling. In our case, Coulomb interaction between the Nb film and in the gold NPs may influence the Cooper pairing. The $T_C$ increase may also be attributed to the coupling between the electron states on the Au NPs and electrons in the Nb, which introduces mixing of states without significant charge transfer. Such mixing may give rise to the zero bias anomalies observed in our tunneling spectra.

This approach is related to the Ginsburg prediction [23], which was proposed for geometry similar to ours. The theory explaining this mechanism is outlined in the supplementary material.

In conclusion, we experimentally measure an enhancement of the critical temperature, which occurs when gold NPs are linked chemically to a superconducting Nb thin film. The observed effect depends on the NPs' size, the length of the organic linkers, and the Nb film thickness. Tunneling spectra acquired on the linked NPs, below $T_C$, exhibited typically zero-bias peaks and occasionally, for the shorter linkers only, also induced proximity gaps. There are several mechanisms that may explain the observations, among them a pairing mechanism that involves coupling between electrons at the gold NPs and on the Nb film, mediated by the vibrations of the organic linker molecule. However, the aforementioned mechanisms do not explicitly account for the peak in the density of states observed in the tunneling spectra.

**Acknowledgments**

Y.P. acknowledges the support of the Peter Brojde Center. R.N. and B.S. acknowledge support of the Israel Science Foundation. O.M. thanks support of DIP 563363, andBSF 2008085. N.K. acknowledges support of ISF 1835/07 and BSF 2008438.

**References**

[1] P. G. D. Gennes, *Superconductivity Of Metals And Alloys*, New ed of 2 Revised ed (Westview Press, 1999).
[2] A. Frydman and Z. Ovadyahu, Europhys. Lett. **33**, 217-222 (1996).
[3] A. Y. Kasumov, M. Kociak, S. Guéron, B. Reulet, V. T. Volkov, D. V. Klinov, and H. Bouchiat, Science **291**, 280 -282 (2001).
[4] S. De Franceschi, L. Kouwenhoven, C. Schonenberger, and W. Wernsdorfer, Nat Nano **5**, 703-711 (2010).
[5] O. Bourgeois, A. Frydman, and R. C. Dynes, Phys. Rev. Lett **88**, 186403 (2002).
[6] O. Yuli, I. Asulin, O. Millo, D. Orgad, L. Iomin, and G. Koren, Phys. Rev. Lett. **101**, 057005 (2008).
[7] M. Brust, D. Bethell, C. J. Kiely, and D. J. Schiffrin, Langmuir **14**, 5425-5429 (1998).
[8] E. Berg, D. Orgad, and S. A. Kivelson, Phys. Rev. B **78**, 094509 (2008).
[9] T. Kommers and J. Clarke, Physical Review Letters **38**, 1091-1094 (1977).
[10] M. G. Blamire, E. C. G. Kirk, J. E. Evetts, and T. M. Klapwijk, Phys. Rev. Lett. **66**, 220 (1991).
[11] S. Yochelis, E. Katzir, Y. Kalcheim, O. Millo, and Y. Paltiel, Submitted for Publication (2011).
[12] D. Shvarts, M. Hazani, B. Y. Shapiro, G. Leitus, V. Sidorov, and R. Naaman, Europhys. Lett. **72**, 465-471 (2005).
[13] E. Katzir, S. Yochelis, and Y. Paltiel, Appl. Phys. Lett. **98**, 223306-3 (2011).
[14] A. A. Abrikosov, Soviet Physics JETP **5**, 1174-1182 (1957).
[15] M. W. Coffey and J. R. Clem, Phys. Rev. Lett. **67**, 386 (1991).
[16] A. E. Hanna and M. Tinkham, Phys. Rev. B **44**, 5919 (1991).
[17] V. Narayanamurti, J. P. Garno, and R. C. Dynes, Physical Review Letters **41**, 1509-1512 (1978).
[18] P. L. Richards and M. Tinkham, Phys. Rev. **119**, 575 (1960).
[19] I. S. Burmistrov, I. V. Gornyi, and A. D. Mirlin, arXiv:1102.3323v1 (2011).
[20] J. Noffsinger and M. L. Cohen, Phys. Rev. B **81**, 214519 (2010).
[21] A. J. Leggett, Phys. Rev. Lett. **83**, 392 (1999).
[22] A. J. Leggett, J Supercond **19**, 187-192 (2006).
[23] V. L. Ginzburg, Sov. Phys. Usp. **19**, 174-179 (1976).

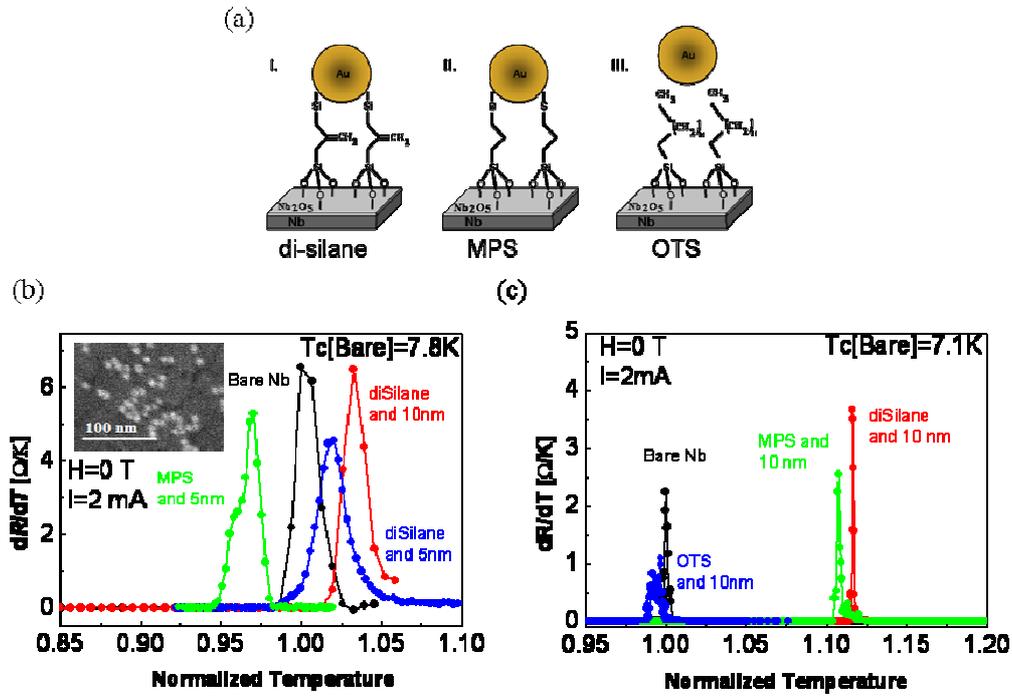

**Figure 1**: (a) Schematics of the samples: self-assembled monolayers of three types of molecules adsorbed on top of Nb films. (I) 3-methylpropane bis-trichlorosilane, (di-silane) which forms a 3-nm-thick layer (II) mercapto propyl silane (MPS), forming a 0.5-nm-thick monolayer, and (III) Tricloro(octadecyl)Silane (OTS) forming a 3 nm thick layer. The Au nano-particles were chemically linked only to the first two types of molecules. (b) The differential resistance as a function of the normalized temperature. The temperature is normalized to $T_C$ of the corresponding bare evaporated Nb film (Black). The blue and red curves show results obtained for samples with 5 and 10 nm Au NPs, respectively, linked to the Nb film via di-silane molecules. In green, data for 5 nm NPs linked by the MPS organic monolayer. The inset shows a SEM image of the Au NPs adsorbed on top of the Nb film. (c) 50nm sputtered Nb film coated with 10-nm NPs linked by the di-silane and MPS organic layer showing about 10% increase in $T_C$. Very small changes in the critical temperature are observed when using non-linking OTS molecules.

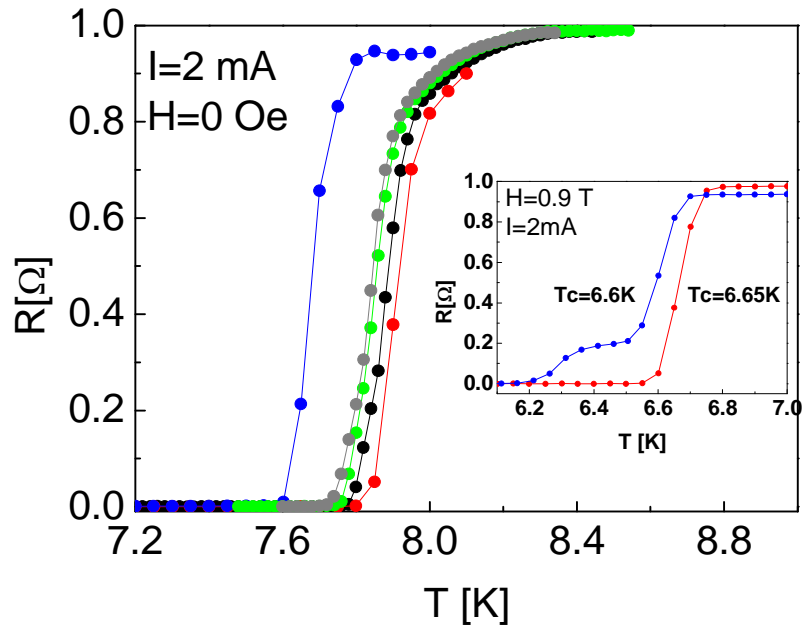

**Figure 2:** R(T) characteristics for 150nm Nb film: Blue- bare Nb substrate, Red- after di-silane adsorption and 10-nm Au NPs deposition. R(T) curves were then measured for the same sample, after partial removal of the organic layer and the Au NPs by plasma. Black- after 10 minutes plasma cleaning, Green- after 20 minutes plasma cleaning and Gray- after 60 minutes plasma cleaning. The data were obtained by applying a current of 2 mA. No effect of the plasma cleaning was seen on the bare Nb film. Inset: R(T) curves before and after the Au NPs attachment measured at 9000 G, close to the matching field, showing smaller $T_C$ enhancement.

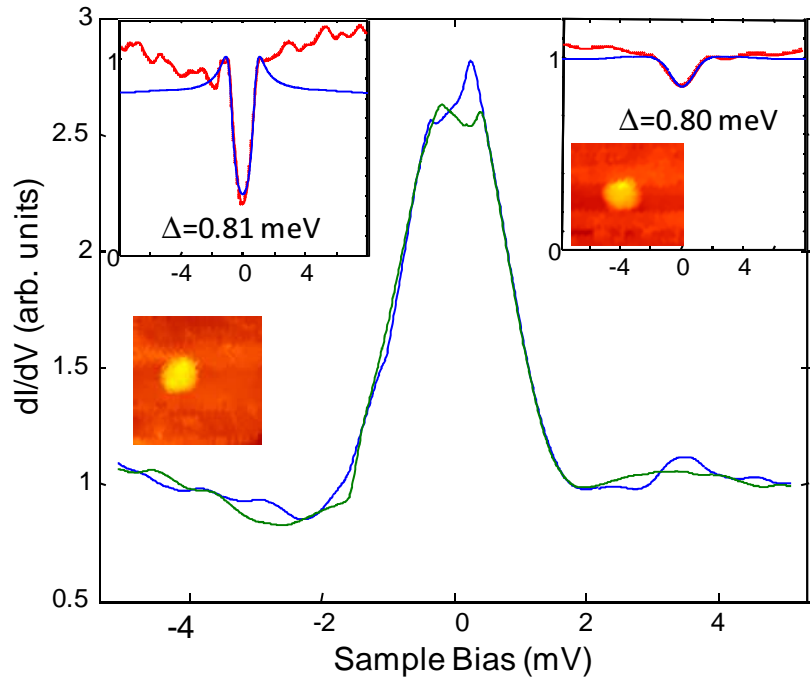

**Figure 3:** (color online) Tunneling spectra acquired at 4.2 K on top of a single Au NP of 10nm diameter (shown to the left) linked to a 50 nm Nb film via di-silane molecules, exhibiting the main spectral feature observed – a zero bias peak. The top right inset presents a spectrum (red line) acquired on a gold NP attached to MPS (shown in the inset), exhibiting a proximity-induced gap, while the spectrum presented in the left inset was taken on the MPS covered Nb surface. The blue lines are fits to the Dynes function (see text), with the indicated gaps. The STM images were taken with set values V = 100 mV and I = 0.1 nA, while the spectra were measured with set voltage of 5 mV. The topographic images are 50nm wide.